\documentclass[reprint, aps, prd, twocolumn, nofootinbib]{revtex4}  %
\usepackage{graphicx}%
\usepackage{dcolumn}%
\usepackage{bm}%
\usepackage[linktocpage,colorlinks,pdfusetitle,allcolors=blue]{hyperref}
\usepackage{amsmath, amssymb}
\usepackage{siunitx}
\usepackage{pgfplots}
\usepackage{tikz}
\usepackage{mathrsfs}
\usepackage[caption=false]{subfig}

\usepackage[cal=boondox]{mathalfa}

\pgfplotsset{compat=newest}
\DeclareSIUnit[]\admmass{\text{\ensuremath{M}}}

\usetikzlibrary{decorations.markings, decorations.pathmorphing}
\tikzset{
  set arrow inside/.code={\pgfqkeys{/tikz/arrow inside}{#1}},
  set arrow inside={end/.initial=>, opt/.initial=},
  /pgf/decoration/Mark/.style={
    mark/.expanded=at position #1 with
    {
      \noexpand\arrow[\pgfkeysvalueof{/tikz/arrow inside/opt}]{\pgfkeysvalueof{/tikz/arrow inside/end}}
    }
  },
  arrow inside/.style 2 args={
    set arrow inside={#1},
    postaction={
      decorate,decoration={
        markings,Mark/.list={#2}
      }
    }
  },
}

\renewcommand{\d}[1]{\ensuremath{\operatorname{d}\!{#1}}}

\begin{document}

\title{Not all spacetime coordinates for general-relativistic ray
  tracing are created equal}

\author{Gabriele Bozzola}
\email{gabrielebozzola@arizona.edu}
\affiliation{Department of Astronomy, University of Arizona, Tucson, AZ, USA}

\author{Chi-kwan Chan}
\email{chanc@arizona.edu}
\affiliation{Department of Astronomy, University of Arizona, Tucson, AZ, USA}
\affiliation{Data Science Institute, University of Arizona, Tucson, AZ, USA}
\affiliation{Program in Applied Mathematics, University of Arizona, Tucson, AZ, USA}

\author{Vasileios Paschalidis}
\email{vpaschal@arizona.edu}
\affiliation{Department of Astronomy, University of Arizona, Tucson, AZ, USA}
\affiliation{Department of Physics, University of Arizona, Tucson, AZ, USA}

\date{\today}

\begin{abstract}
  Models for the observational appearance of astrophysical black holes rely
  critically on accurate general-relativistic ray tracing and radiation
  transport to compute the intensity measured by a distant observer. In this
  paper, we illustrate how the choice of coordinates and initial conditions
  affect this process. In particular, we show that propagating rays from the
  camera to the source leads to different solutions if the spatial part of the
  momentum of the photon points towards the horizon or away from it. In doing
  this, we also show that coordinates that are well suited for numerical
  General-Relativistic MagnetoHydroDynamic (GRMHD) simulations are typically not
  optimal for generic ray tracing. We discuss the implications for black-hole
  images and show that radiation transport in optimal and non-optimal spacetime
  coordinates lead to the same images up to numerical errors and algorithmic
  choices.
\end{abstract}

\maketitle

\section{Introduction}%
\label{sec:introduction}

The successful observations at the scale of black-hole horizons by the Event
Horizon Telescope (EHT) Collaboration~\cite{EHT1, EHT2, EHT3, EHT4, EHT5, EHT6,
  EHT7, EHT8, EHTsgr1, EHTsgr2, EHTsgr3, EHTsgr4, EHTsgr5, EHTsgr6} highlight
the need for accurate general-relativistic ray tracing and radiation transfer.
Ray tracing consists of finding the path that photons take from the hot plasma
around black holes to our camera. In general relativity, we assume that these
photons do not back-react onto the spacetime and are not quantum in nature, so
the problem is equivalent to finding null geodesics (hereafter,
rays).\footnote{Understanding the geodesic structure is one of the primary tools
  in investigating spacetimes. Hence, the literature on the topic of general
  relativistic ray tracing is massive and spans several decades. At this point,
  there is a large number of codes available for ray tracing, several of which
  are public (for a non-comprehensive list see, e.g.~\cite{Gold2020}).} Once the
path of a ray is known, one has to perform radiation transfer. This consists of
computing specific intensity and optical depth along the ray, assuming some
models for how they change due to the local thermodynamical properties of the
plasma.\footnote{A different approach, e.g., Monte Carlo radiative transfer
  \cite{2009ApJS..184..387D}, might be preferred if the system under
  consideration is dominated by scattering.} Ray tracing and radiation transfer
are critical for the calibration, modeling, and interpretation of EHT results
(e.g.~\cite{Gralla2019, EHT5, EHTsgr5, Younsi2021}). They are also employed in
accurate modeling of electromagnetic radiation from other compact objects such
as stellar mass black holes~\cite[e.g.,][]{2016ApJ...819...48S} and neutron
stars~\cite[e.g.,][]{2012ApJ...753..175B, 2019ApJ...872..162B}.

In this paper, we highlight how ray tracing is affected by choices of
coordinate systems, initial conditions, and direction of integration (we will make this
and other similar statements more precise below). First, we show that ray
tracing is generally not invariant with respect to as time reversal:
choosing our initial condition at the camera,
integrating a geodesic between the source and the camera leads to two different
results if the integration is performed forward (with the photon spatial
momentum pointing towards the source) or backward in time (with the photon
spatial momentum pointing away from the source). This is yet another example of
how black holes defy our intuition. In non-relativistic ray tracing (e.g., for
computer graphics, or astronomy), there exist only one curve that connects a
pixel of the camera to the source and this curve can be found by shooting rays
from the camera to the source. On this curve, information can freely flow in both
the directions. This is no longer true in general relativity (unless both
spacetime and the matter are also invariant with respect to time reversal).

Second, we discuss how some coordinate systems are better suited for ray tracing
than others. In particular, coordinates used in stationary spacetime
General-Relativistic MagnetoHydroDynamic (GRMHD) simulations are typically not
an optimal choice for ray tracing. The reason for this is that coordinates for
GRMHD simulations are designed to let information flow towards the horizon, but
ray tracing amounts to collecting information \emph{from} the horizon, which is
the opposite problem. Choice of non-optimal coordinates leads to numerical
problems, and while gauge-invariant procedures ought to yield
gauge-invariant results, in general this is not the case due to numerical
considerations. We will show an explicit example in which the numerical error
diverges due to the impossibility for numerical algorithms to properly resolve
the geodesic. On the other hand, other coordinates lead to well behaved
geodesics that numerical schemes can easily integrate. We will also discuss that
in general the error in computing black-hole images through general-relativistic
radiation transfer is small.

The structure of the paper is as follows. In Section~\ref{sec:setup}, we review
the fundamental tools that we need for the rest of the study: ray tracing and
the Kerr-Schild (KS) coordinates. In Section~\ref{sec:schwarzschild}, we
describe our findings focusing on the non-spinning case. This case is insightful
as it can be more easily visualized and understood in terms of conformal
diagrams\footnote{We would like to thank Erik Wessel for suggesting this way of
  looking at the problem.}. Next, in Section~\ref{sec:kerr} we discuss the
rotating case and the implications for black-hole images. Finally, we collect
our conclusions in Section~\ref{sec:conclusions}. We use units with $G=c=1$,
where $G$ is the gravitational constant, and $c$ the speed of light in vacuum.
We use the same conventions as~\cite{MTW1973}: the signature of the metric is
$(-, +, +, +)$, indices go from 0 to 3 (with 0 being the time component), and
employ the Einstein summation convention for repeated indices.

\section{Setup}%
\label{sec:setup}

\subsection{Ray tracing}%
\label{sec:ray-tracing}

General-relativistic ray tracing requires the solution of the geodesic
equation for photons that reach an observer far away from the source. Let
$x^{\mu}(\lambda)$ be the null geodesic we want to reconstruct, we have that
\begin{equation}
  \label{eq:geo-eq-chri}
  \frac{\d{}^{2} x^{\mu}}{\d \lambda^{2}} = - \Gamma^{\mu}_{\alpha\beta} \frac{\d x^{\alpha}}{\d \lambda}\frac{\d x^{\beta}}{\d \lambda} \,,
\end{equation}
where $\Gamma^{\mu}_{\alpha\beta}$ are the Christoffel symbols and $\lambda$ is the affine parameter. We
say that a geodesic $x^{\mu}(\lambda)$ is affinely parametrized when it satisfies
Equation~\eqref{eq:geo-eq-chri}. Christoffel symbols are symmetric in the lower
indices and can be computed from the metric $g_{\alpha\beta}$ as
\begin{equation}
  \label{eq:christ}
  \Gamma^{\mu}_{\alpha\beta} = \frac{1}{2} g^{\mu\gamma} \left(\partial_{\alpha} g_{\beta\gamma} + \partial_{\beta} g_{\gamma\alpha} -\partial_{\gamma} g_{\alpha\beta}\,. \right)
\end{equation}
This second order equation can be cast into a system of first order ones,
\begin{subequations}
  \begin{align}
    \label{eq:geo-eq-chri-sys}
    \frac{\d x^{\mu}}{\d \lambda} &= k^{\mu}\,, \\
    \frac{\d k^{\mu}}{\d \lambda} &= - \Gamma^{\mu}_{\alpha\beta} k^{\alpha} k^{\beta} \,.
  \end{align}
\end{subequations}
$k^{\mu}$ is the vector tangent to the geodesic, and it is a null vector, so it
has to be that $k^{\mu} k_{\mu} = 0$ along the null geodesic. We will refer to this
as the constraint of the problem.

Equations~\eqref{eq:geo-eq-chri-sys} have to be supplemented with initial
conditions for $x^\mu$ and $k^\mu$. For this, we follow~\cite{Psaltis2012, Chan2013}
in setting up a Cartesian grid with coordinates $\alpha$ and $\beta$ (the pixels of our
camera) perpendicular to the line of sight at a large Euclidean distance $d$
from the source and inclination from the pole $\mathcal{i}$ and azimuthal angle
$\mathcal{j}$. As commonly done in the field, the inclination $\mathcal{i}$ is
such that $\mathcal{i} = 90{}^{\circ}$ corresponds to the equatorial plane. For a
given pixel, we choose the spatial components of $k^{\mu}$ such that they are
perpendicular to the image plane, and use the condition that $k^{\mu} k_{\mu} = 0$
to normalize $k^{t}$ to 1. Setting $k^{t} = 1$ means that we can interpret the
affine parameter $\lambda$ as the coordinate time at the camera, and that time flows
forward for increasing values of $\lambda$. With this choice, we will say that we
integrate forward (backward) in time when we integrate with increasing
(decreasing) values of $\lambda$. We will also say that the photon points towards
(away from) the source if the spatial part of $k^{\mu}$ is outgoing from (ingoing to)
the camera. In some codes, all these quantities are associated to integrals of
motion (energy, angular momentum, and Carter constant).

We implement this scheme in a new code, \texttt{kRay}~\cite{kRay}, where we
solve Equations~\eqref{eq:geo-eq-chri-sys} numerically with the LSODA
solver~\cite{LSODA} of ODEPACK~\cite{hindmarsh1982odepack} through the SciPy
interface~\cite{SciPy}. The LSODA solver is accurate and robust, it has adaptive
stepping and automatic stiffness detection (so that an implicit scheme is used
when necessary), and can provide high-order interpolating functions. All the
numerical integrations presented in this paper are performed setting \num{e-14}
as absolute and relative tolerance, and all the calculations are in double
precision.

\subsection{Kerr-Schild coordinates}%
\label{sec:kerr-schild-coord}

The Kerr spacetime describes a black hole with mass $M$ and angular momentum
$Ma$. In Cartesian Kerr-Schild (KS) coordinates $(t,x,y,z)$, this is defined by
the metric
\begin{equation}
  \label{eq:kerr-schild}
  g_{\alpha\beta} = \eta_{\alpha\beta} + fl_{\alpha} l_{\beta}\,,
\end{equation}
with $\eta_{\alpha\beta}$ being the flat-spacetime Minkowski metric in Cartesian coordinates, and
\begin{align}
  f &= \frac{2 M r^{3}}{r^{4} + a^{2} z^{2}} \,, \label{eq:ks-f} \\
  l_{\alpha} &= \left(\pm 1, \frac{rx + ay}{r^{2} + a^{2}}, \frac{ry - az}{r^{2} + a^{2}}, \frac{z}{r} \right)  \label{eq:ks-la}\,,
\end{align}
where the plus and minus signs in the time component correspond to the
\emph{ingoing} and \emph{outgoing} Kerr-Schild coordinates, respectively, and
$r$ is implicitly defined by
\begin{equation}
  \label{eq:r-kerr-schild}
  r^{2} + a^{2}\left(1 - \frac{z^{2}}{r^{2}}\right) = x^{2} + y^{2} + z^{2}\,.
\end{equation}
In Cartesian Kerr-Schild, the horizon is a coordinate ellipsoid described by the
equation
\begin{equation}
  \label{eq:rh-kerr-schild}
  \frac{x^{2}}{R_{H}^{2} + a^{2}} + \frac{y^{2}}{R_{H}^{2} + a^{2}} + \frac{z^{2}}{R_{H}^{2}} = 1\,,
\end{equation}
with $R_{H} = M + \sqrt{M^{2} - a^{2}}$. These coordinates are well defined
everywhere (except at $r=0$), they are horizon-penetrating, meaning that all the
important fields (metric, Christoffel symbols, et cetera) are well-behaved on
the horizon. For this reason, ingoing Kerr-Schild (or slight variations of them)
are ubiquitously employed in numerical simulations of accretion flows \cite[see,
  e.g.,][]{2003ApJ...589..444G}. We will use Kerr-Schild
coordinates to discuss our findings but the issues presented are general
features of general-relativistic ray tracing. Other coordinates that are
commonly used are Boyer-Lindquist, which are not horizon penetrating for either
ingoing or outgoing causal curves. Many of the early ray tracing codes
(e.g.,~\cite{1992MNRAS.259..569K, 1993A&A...272..355V, 1994ApJ...421...46R,
  2009ApJ...696.1616D, 2012ApJ...745....1P, 2012ApJ...753..175B,
  2013ApJ...777...13C}) use the Boyer-Lindquist coordinates to take advantage of
the symmetry of the spacetime and to reduce the computation
cost.\footnote{See~\cite{Christian2020} for a discussion and comparison with the
  Boyer-Lindquist coordinates.} However, supporting arbitrary coordinate
systems~(e.g.,~\cite{2009ApJS..184..387D}) or adopting Kerr-Schild-like
coordinates~(e.g.,~\cite{2018ApJ...867...59C}) is preferred for interacting with
GRMHD simulations. Note that even if the geodesics of Kerr spacetimes are
integrable and closed forms exist (e.g.~\cite{2009ApJ...696.1616D, Gralla2020}),
as photon positions and momenta ($\propto k^\mu$) are frequently needed to
sample the plasma along geodesics,
in practice, it is often simpler and faster to perform the integration
numerically, which is what most codes do.

\section{Different coordinates lead to different results}%
\label{sec:schwarzschild}

In this Section, we identify and discuss two features of the ray-tracing scheme
previously described. First, ray tracing leads to different results if the
integration is performed forward in time with photon pointing towards the source
or backward with photon pointing away (as defined in
Section~\ref{sec:ray-tracing}). Second, the choice of coordinates and initial
conditions dictate which geodesics can be properly reconstructed, so some
coordinate systems are better suited for ray tracing than others. We will
present these results by considering a Kerr black hole with $a=0$ (the
Schwarzschild spacetime) in ingoing Kerr-Schild coordinates. In this case, we
can understand most of the features we want to present using accessible
equations and diagrams.

The problems we want to discuss already arise in one of the simplest cases: ray
tracing a photon on the $x$ axis (because of rotational symmetry, this is
equivalent to any purely radial integrations). Let us focus on the $y=z=0$ line
with $x > 0$ (where $r=x$), we have that
\begin{equation}
  \label{eq:schwarzschild-kerr-schild}
  \d s^{2} = -\left(1 - \frac{2M}{x}\right)\d t^{2} + \frac{4M}{x} \d t \d x
  +\left(1 + \frac{2M}{x}\right) \d x^{2} \,,
\end{equation}
where $\d s^{2} = g_{\mu\nu} \d x^{\mu} \d x^{\nu}$ is the element of proper spacetime
length. We can understand most of the geodesic properties by looking at the null
cones on this line. This is done by setting $\d s^{2} = 0$, defining
$\dot{x} = {\d x}\slash{\d t}$:
\begin{equation}
  \label{eq:null-equation}
   \left(1 + \cfrac{2M}{x} \right){\dot{x}}^{2} + \frac{4M}{x} \dot{x} - \left(1 - \cfrac{2M}{x} \right)= 0\,.
\end{equation}
Solving this, we find that
\begin{equation}
  \label{eq:null-solution-eq}
  \frac{\d x}{\d t} = \frac{-\cfrac{2M}{x} \pm \sqrt{{\left(\cfrac{2M}{x}\right)}^{2} + \left(1 - \cfrac{2M}{x} \right)\left(1 + \cfrac{2M}{x} \right) }}{1 + \cfrac{2 M}{x}}\,,
\end{equation}
simplify,
\begin{equation}
  \label{eq:two-null-solutions}
  \frac{\d x}{\d t} = - 1 \text{ or } \frac{\d x}{\d t} = \frac{x - 2 M}{x + 2 M}\,,
\end{equation}
which, after integration, lead to
\begin{equation}
  \label{eq:two-null-solutions-integrated}
  t(x) = - x + C_{1} \text{ and } t(x) = x + 4 M \ln (x - 2M) + C_{2}\,,
\end{equation}
with $C_{1}, C_{2}$ integration constants with units of $\si{\admmass}$.

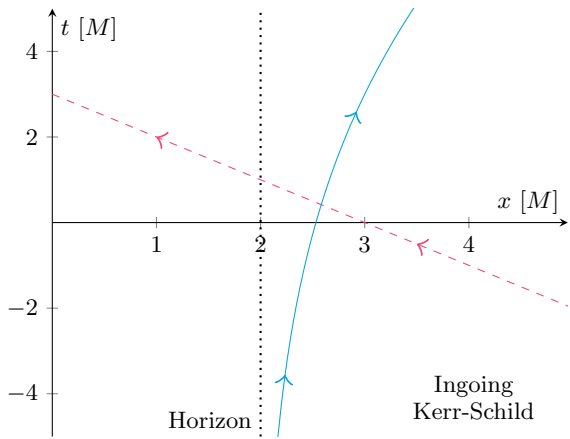
\begin{figure}[htbp]
  \centering
  \begin{tikzpicture}
\pgfplotsset{
  BaseStyle/.style={
    tick pos=left,
    ytick style={color=black},
    xtick style={color=black},
    grid style={line width=.2pt, draw=gray!30},
    legend cell align={left},
    legend style={line width=0.25pt, cells={align=left}},
    xminorgrids,
    xmajorgrids,
    yminorgrids,
    ymajorgrids,
  },
}

\definecolor{color0}{HTML}{06a5d6}
\definecolor{color1}{HTML}{ef476f}
\definecolor{color2}{HTML}{2cd360}
\definecolor{color3}{HTML}{118ab2}
\definecolor{color4}{HTML}{ffd166}

\pgfplotsset{
  ThickStyle0/.style={
    color0,
    dashed,
    very thick
  },
}

\pgfplotsset{
  ThickStyle1/.style={
    color1,
    very thick,
    dotted
  },
}

\pgfplotsset{
  ThickStyle2/.style={
    very thick,
    color2,
    dash dot
  },
}

\pgfplotsset{
  ThickStyle3/.style={
    very thick,
    color3,
    loosely dashed
  },
}

\pgfplotsset{
  ThickStyle4/.style={
    very thick,
    color4,
    loosely dotted
  },
}
     \begin{axis}[
      axis lines = middle,
      ymin=-5, ymax=5,
      xmax=4.95,
      xlabel={$x~[M]$},
      ylabel={$t~[M]$},
      ]
      \addplot[domain=5:0, color1, dashed] {-x + 3} [arrow inside={opt={scale=2}}{0.3, 0.8}];
      \addplot[domain=2.05:5, samples=100, color0] {x + 4 * ln(x - 2)}[arrow inside={opt={scale=2}}{0.3, 0.6}];
      \draw[dotted, thick] (axis cs:2, -5) -- (axis cs:2, 5);
      \node[anchor=south east] at (axis cs:2, -5) {Horizon};

      \node[align=center, text width = 2cm, anchor=south east] at (rel axis cs:0.98, 0.02) {Ingoing Kerr-Schild};
    \end{axis}
  \end{tikzpicture}
  \caption{Example of null geodesics in ingoing Kerr-Schild coordinates. Ingoing
    geodesics (arrows pointing towards $x=0$) are horizon-penetrating and
    well-behaved. Outgoing (arrows pointing towards $x=+\infty$) geodesics are
    singular at the horizon. The choice of the direction of integration (forward
    or backward in time) determines which of the two families is being solved
    for. This shows that it is impossible for observers at infinity to collect
    information coming from inside the horizon.}%
  \label{fig:outgoing}
\end{figure}

Equation~\eqref{eq:two-null-solutions-integrated} describes the null radial
geodesics and establishes the existence of two families of solutions, ingoing
($\d x/\d t<0$) and outgoing photons ($\d x/\d t>0$). Figure~\ref{fig:outgoing}
shows one example from each of these families on a spacetime diagram, where the
ingoing ray is red dashed and the outgoing one is solid blue. Fixing $k^{t} = 1$
and assuming that integration starts from a camera far away (as in
Section~\ref{sec:ray-tracing}), the integration will select one of the two
solutions depending on the initial conditions: when the geodesic is
integrated forward in $\lambda$ (i.e., with increasing values of $\lambda$) with the spatial
part of $k^{\mu}$ pointing towards the source, the solution will be the ingoing
one. The other one is selected when the geodesic is integrated backward in $\lambda$
(i.e., with decreasing values of $\lambda$), with the spatial part of $k^{\mu}$ pointing
away from the horizon (while still assuming $k^{t} = 1$). The existence of two
distinct families demonstrates an important feature of general relativistic ray
tracing: integrating photons backward in time is not the same as reversing their
initial spatial momentum and integrating forward. Compare this with
non-relativistic ray tracing where one can shoot a ray and traverse it in both
directions.\footnote{Assuming time-symmetric matter, integrating towards the
  horizon in ingoing coordinates is equivalent to integrating backwards in
  outgoing ones. This is not generally true for other spacetimes (including
  Kerr).}

\begin{figure}[htbp]
  \centering
  \includegraphics[]{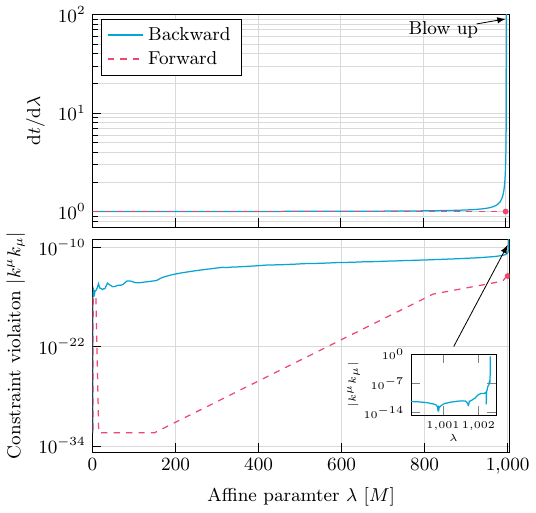}
  \caption{(Top.) Time component of the photon four-velocity for a ray
    integrated backwards in time moving towards the camera and one integrated
    forward in time moving towards the horizon. (For the former, we redefined
    $\lambda \to -\lambda$ to enable the comparison). The filled circle indicates when the
    photon crosses the horizon. In the case of backward propagation, the photon
    never does so. The fast growth in $\d t \slash \d \lambda$ leads to a growth in the
    constraint violation as well, as shown in the bottom panel. (Bottom.) The
    constraint grows to arbitrary high values. The solution for the forward
    integration is constant, so the algorithm can take large steps (the entire
    solution took only 11 steps) and keep the error down. The filled circle
    indicates when the photon crosses the horizon. In the case of backward
    propagation, the photon never does so and one has to impose some artificial
    prescription to end the integration.}%
  \label{fig:kt-schw}
\end{figure}

Equation~\eqref{eq:two-null-solutions-integrated} shows that outgoing rays
diverge exponentially at the finite radius $r= \SI{2}{\admmass}$, while ingoing
rays are always well behaved. For outgoing photons near the horizon, $\dot{x}
\to 0$, which means that it takes an infinite amount of coordinate time $t$ to
make any infinitesimal step in $x$. In standard affine parametrizations, this
manifests itself in ${\d t}\slash{\d \lambda}$ diverging near the horizon, as we
explicitly show in the appendix. Numerical schemes, even sophisticated ones,
cannot accurately reconstruct this behavior. For instance, methods with adaptive
timestepping will want to take an infinitesimally small step to keep the error
under control. However, quickly the step becomes smaller than the finite
precision of the machine and the integration cannot continue. This behavior is
shown in Figure~\ref{fig:kt-schw}, which depicts $\d t \slash \d \lambda$ on the
top panel and the growth of the constraint (deviation of $\lvert k^{\mu} k_{\mu}
\rvert$ from zero) in the second.

For outgoing rays, the photon will never touch the horizon in finite time given
that $\dot{x} \to 0$.\footnote{Note, this is a general feature of event horizons
  and it does not depend on the details of the coordinate chosen. Event horizon
  are defined as the boundary of causal past of future null infinity
  $\cal{I}^{+}$~\cite{Wald1984}, so no causal curve can connect events from
  inside the horizon to our camera. If the opposite were true, we would be able
  to see inside black holes.} Hence, the most natural termination condition for
the numerical integration (the photon crossing the horizon) will never occur. As
a result, one has to impose a different, artificial termination condition (e.g.,
stop the integration at some finite distance from the horizon).
Figure~\ref{fig:kt-schw} shows that it is numerically impossible to reach
distances that are arbitrarily close to the horizon and the result of the
integration will depend on a prescribed stopping condition. On the other hand,
ingoing photons are perfectly well behaved and cross the horizon in finite time.
In fact, the solution for ingoing photon is constant, so numerical schemes can
capture this easily and accurately, and the entire solution requires a handful
of steps to achieve an accuracy of better than \num{e-14}. In addition to
keeping the error small, no artificial stopping condition has to be prescribed
to terminate the integration. In other words, we can perfectly reconstruct the
geodesics of ingoing photons, but we will always introduce errors in computing
the ones for outgoing rays.

Figure~\ref{fig:kt-schw} seems to suggest that ray tracing with integrated rays
backward from the camera to the horizon is always bound to incur in significant
numerical problems. We now show that this is purely a coordinate
effect.\footnote{Similar conclusions were obtained in~\cite{Pihajoki2018}.
  Another example of the use of outgoing Kerr-Schild coordinates in the context
  of ray tracing is~\cite{Riazuelo2020}.} To do so, we move to \emph{outgoing}
Kerr-Schild coordinates by choosing the minus sign in Equation~\eqref{eq:ks-f}.
The transformation only changes the sign in front of the first $2M\slash x$ at the
numerator of Equation~\eqref{eq:null-solution-eq}, so we the null cones satisfy
\begin{equation}
  \label{eq:null-solution-fixed}
  \frac{\d x}{\d t} = \frac{\frac{2M}{x} \pm 1}{\frac{2M}{x} + 1}\,.
\end{equation}
The two solutions are
\begin{equation}
  \label{eq:two-null-solutions-fixed}
  \frac{\d x}{\d t} = 1 \qquad \frac{\d x}{\d t} = \frac{2 M - x}{2 M + x}\,.
\end{equation}
We can integrate these to obtain
\begin{equation}
  \label{eq:two-null-solutions-integrated-fixed}
  t(x) = x + C_{3} \qquad t(x) = - x - 4 M \ln (2M - x) + C_{4}\,,
\end{equation}
with $C_{3}$, $C_{4}$ integration constants with units of $\si{\admmass}$. This
change in coordinates completely flips the situation presented above. In the new
coordinates, the outgoing null rays are well-behaved and the ingoing ones are
not. For static spacetimes, we can perform an isometric identification to map
the outgoing geodesic in the outgoing coordinates to the ingoing geodesic in the
ingoing coordinates. Therefore, only in these spacetimes (which include
Schwarzschild but not Kerr), it makes sense to integrate towards the source.
Note that is no longer true in presence of non static sources. In other words,
for Schwarzschild, moving from ingoing Kerr-Schild to outgoing Kerr-Schild is
equivalent to reversing the integration of the photon. So, we can read
Figure~\ref{fig:kt-schw} as comparing the two coordinate systems for outgoing
rays (with red dashed line showing the result in outgoing Kerr-Schild and the
solid blue line in ingoing ones).

This simple case shows that if we want to perform accurate ray tracing
integrating backward in time, the outgoing Kerr-Schild coordinates are a
superior choice. This conclusion is ultimately a statement about the causal
structure of the coordinates and does not depend on the specific example
considered. Ingoing Kerr-Schild coordinates are a good choice for numerical
simulations because matter can easily flow inside the horizon, but they are not
suitable for ray tracing, where we want to propagate information from regions
near the horizon. In Section~\ref{sec:poss-work-integr}, we discuss a possible
way to implement stable integrations in ingoing Kerr-Schild.

\onecolumngrid{}

\begin{figure}[htbp]
  \centering
  \includegraphics[]{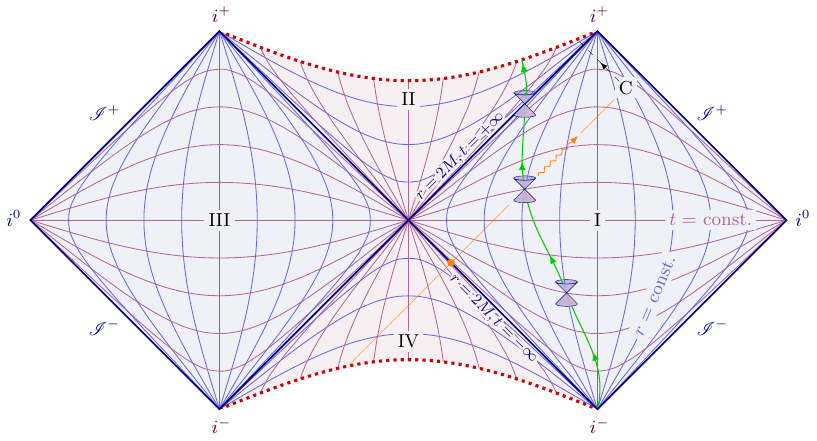}
  \caption{Conformal diagram for a maximally-extended Schwarzschild spacetime
    (the angular dimensions are suppressed). Null geodesics are inclined with a
    45${}^\circ$ angle, with future directed ones moving towards
    $\mathscr{I}^{+}$.  The ingoing Kerr-Schild coordinates only cover regions I
    and II, the outgoing ones region I and IV\@. Our universe is region
    I. Constant time and radius curves are hyperbolas (purple and blue
    respectively). We place our camera C at a large separation from the horizon
    and trace back the emission along the orange line. For the ingoing
    Kerr-Schild coordinates, the propagation hits a coordinate singularity at
    $r=\SI{2}{\admmass}$ at the past horizon (orange dot) because that part of
    the spacetime is not mapped in these coordinates, whereas integration
    proceeds uninhibited until the physical singularity in the outgoing
    coordinates. The diagram also shows how integrating backward and forward in
    time results in two different geodesics: the orange one (backward) and the
    black dashed one (forward). The world line of the fluid element from which
    we collect emission is in green. Emission that is detectable from our
    cameras only comes from particles in region I.  The problem with performing
    ray-tracing in ingoing coordinates is that there is ``coordinate barrier''
    at $r=\SI{2}{\admmass}$ at the past horizon, so numerical algorithms fail.}%
  \label{fig:pen-kru}
\end{figure}

\twocolumngrid{}

Figure~\ref{fig:pen-kru} shows the conformal diagram for maximally extended
Schwarzschild spacetime. This plot, representing null geodesics as lines at
45${}^\circ$ angle, concisely explains all the features that we have described
so far. Our universe is described by region I and our camera C is a time-like
observer lying at a constant large radius. The boundary between region I and II
is known as the future event horizon, and that between region I and IV (the past
event horizon). The green solid line with light cones represents the world line
of a parcel of plasma that will eventually fall into the horizon. Ray tracing
and radiation transfer consists of modeling and collecting all the emission from
these fluid elements.

First, we can see directly that integration backward and forward in time, as
defined earlier, lead to different geodesics (compare the dashed black arrow
line originating from the camera C with the orange solid line). So, when we
perform ray tracing, we have to integrate backwards from the camera C with
momentum pointing away from the horizon. Second, we can understand the relative
performance between the two Kerr-Schild coordinates by looking at which
quadrants they describe. Ingoing Kerr-Schild describes regions I and II, so that
we can follow the trajectory of the fluid element towards the horizon. When we
perform ray tracing along the orange geodesic, we run into a coordinate
singularity near the boundary with region IV\@. The geodesic is not complete and
there is a coordinate barrier on that boundary, which is felt when $r \to
\SI{2}{\admmass}$. On the other hand, outgoing Kerr-Schild describes region I
and IV, so we can reconstruct the geodesic in its entirety and there are no
issues with coordinates.

\subsection{Mitigating numerical instabilities by integrating in coordinate time}%
\label{sec:poss-work-integr}

\begin{figure}[htbp]
  \centering
  \includegraphics[]{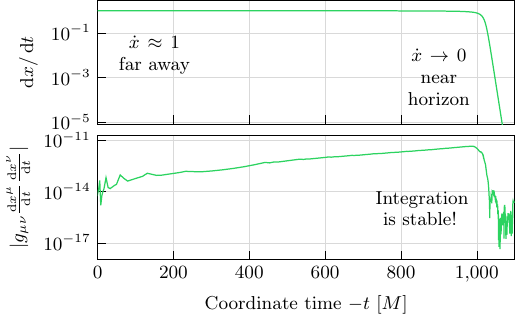}
  \caption{Backward propagation in coordinate time. This scheme does not have
    numerical problems and the integration can be carried on for arbitrarily
    long times. The photon will take an infinite amount of time to reach the
    horizon. $\dot{x}$ drops to zero near unstable part of the integration is
    pushed to infinity. }%
  \label{fig:u2-constraint-coord}
\end{figure}

In Section~\ref{sec:schwarzschild}, we discussed how it is not possible to
achieve long-term stability in numerical integration of null geodesics in
coordinate systems that are not well adapted to the problem. In the case of
Kerr-Schild, the horizon is a coordinate singularity, and the impossibility of
extending some geodesics past it is a pure gauge effect that manifests itself in
the uncontrollable growth of $k^{t}$. A possible way to continue the integration
without running in numerical problems is to perform the geodesic integration in
coordinate time $t$ as opposed to affine parameter $\lambda$. To do so, the equation
that has to be solved is
\begin{equation}
  \label{eq:geo-coord-12}
   \frac{\d{}^{2} x^{\mu}}{\d t^{2}} =  \left(\Gamma^{0}_{\alpha\beta} \frac{\d x^{\mu}}{\d t} - \Gamma^{\mu}_{\alpha\beta}  \right) \frac{\d x^{\alpha}}{\d t} \frac{\d x^{\beta}}{\d t} \,.
\end{equation}
Figure~\ref{fig:u2-constraint-coord} shows that the formulation is stable.

Let us derive Equation~\eqref{eq:geo-coord-12}. Consider a geodesic described by
$x^{\mu}(\lambda) = \left(t(\lambda), x^{i}(\lambda)\right)$ with $i \in \{1, 2, 3\}$ and $\lambda$ is the
affine parameter. Starting from Equation~\eqref{eq:geo-eq-chri},
\begin{equation}
  \label{eq:geo-eq-chri-2}
  \frac{\d{}^{2} x^{\mu}}{\d \lambda^{2}} = - \Gamma^{\mu}_{\alpha\beta} \frac{\d x^{\alpha}}{\d \lambda}\frac{\d x^{\beta}}{\d \lambda} \,,
\end{equation}
we set $\mu=0$ and find
\begin{equation}
  \label{eq:geo-coord-1}
  \frac{\d{}^{2} t}{\d \lambda^{2}} = - \Gamma^{0}_{\alpha\beta} \frac{\d x^{\alpha}}{\d \lambda} \frac{\d x^{\beta}}{\d \lambda}\,,
\end{equation}
Using the chain rule, we have that
\begin{equation}
  \label{eq:geo-coord-2}
  \frac{\d{}^{2} x^{\mu}}{\d \lambda^{2}} =  \frac{\d{}}{\d \lambda} \left( \frac{\d x^{\mu}}{\d \lambda} \right) =
 \frac{\d{}}{\d \lambda} \left( \frac{\d x^{\mu}}{\d t} \frac{\d t}{\d \lambda} \right)\,.
\end{equation}
According to the Leibniz rule and applying again the chain rule on the first term,
we obtain
\begin{equation}
  \label{eq:geo-coord-3}
  \frac{\d{}}{\d \lambda} \left( \frac{\d x^{\mu}}{\d t} \frac{\d t}{\d \lambda} \right) = {\left(\frac{\d t}{\d \lambda}\right)}^{2} \frac{\d{}^{2} x^{\mu}}{\d t^{2}} + \frac{\d{}^{2} t}{\d \lambda^{2}} \frac{\d x^{\mu}}{\d t}\,.
\end{equation}
Using Equation~\eqref{eq:geo-coord-1}, we can write
\begin{align}
  \label{eq:geo-coord-4}
  {\left(\frac{\d t}{\d \lambda}\right)}^{2} \frac{\d{}^{2} x^{\mu}}{\d t^{2}} + \frac{\d{}^{2} t}{\d \lambda^{2}} \frac{\d x^{\mu}}{\d t} = \\
  {\left(\frac{\d t}{\d \lambda}\right)}^{2} \frac{\d{}^{2} x^{\mu}}{\d t^{2}} - \Gamma^{0}_{\alpha\beta} \frac{\d x^{\alpha}}{\d \lambda} \frac{\d x^{\beta}}{\d \lambda} \frac{\d x^{\mu}}{\d t} =\\
  \label{eq:geo-coord-5}
  {\left(\frac{\d t}{\d \lambda}\right)}^{2} \frac{\d{}^{2} x^{\mu}}{\d t^{2}} - \Gamma^{0}_{\alpha\beta}{\left(\frac{\d t}{\d \lambda}\right)}^{2}  \frac{\d x^{\alpha}}{\d t} \frac{\d x^{\beta}}{\d t} \frac{\d x^{\mu}}{\d t}\,,
\end{align}
where we applied the chain rule in the last step. The left-hand-side of
Equation~\eqref{eq:geo-eq-chri-2}  can be substituted with
Equation~\eqref{eq:geo-coord-5}, using the chain rule on that equation one more
time, eliminating ${\left({\d t}\slash{\d \lambda}\right)}^{2}$ and re-arranging terms, we
find
\begin{equation}
  \label{eq:geo-coord-11}
   \frac{\d{}^{2} x^{\mu}}{\d t^{2}} = \left(\Gamma^{0}_{\alpha\beta} \frac{\d x^{\mu}}{\d t} - \Gamma^{\mu}_{\alpha\beta}  \right) \frac{\d x^{\alpha}}{\d t} \frac{\d x^{\beta}}{\d t} \,.
\end{equation}
Note that this equation is well-defined only when ${\d t}\slash{\d \lambda}$ is finite.
Analytically, this condition is always satisfied except on the horizon.

The ultimate reason why this formulation works is because we traded an integration in
a finite time (but that diverges), with one that takes an infinite amount of time
and becomes unstable only for $t \to \infty$. With this formulation, we can integrate
arbitrarily long in the past without running into numerical problems, as shown
in Figure~\ref{fig:u2-constraint-coord}, which reports the constraint violation
for the same setup described in Section~\ref{sec:schwarzschild}. While this
formulation works, calculations with analytical models will produce more
accurate results if using improved coordinate systems described in the main
text. In practice, with an appropriate coordinate
transformation~\cite{Christian2020}, this formulation is also useful for
ray-tracing GRMHD simulations, which typically use ingoing Kerr-Schild and where
a natural cutoff in coordinate time when to stop the integration already exists.

\section{Black hole images}%
\label{sec:kerr}

The high degree of symmetry of the Schwarzschild spacetime allowed us to clearly
analyze the problem and understand what happens in terms of equations and
diagrams. This is no longer possible for most other spacetimes, including Kerr.
Nonetheless, the features described in the previous Section are presented in
those cases as well and when coordinates are not adapted to the problem it is
not possible to fully integrate the geodesics. In this Section, we look at the
more general rotating case and highlight features that only depend on the
coordinates used. Next, we discuss gauge-invariant observables, like black-hole
images and shadows.

\begin{figure}[htbp]
  \centering
  \includegraphics[]{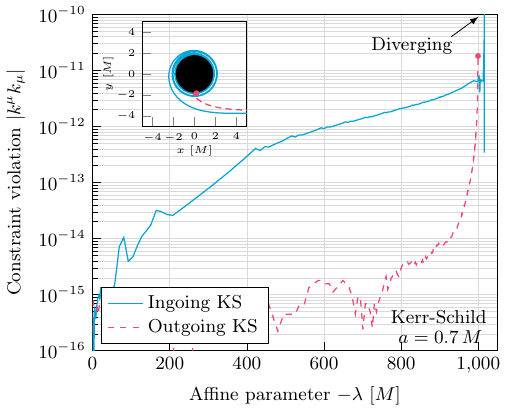}%
  \caption{Numerical violation of the null condition for a null geodesic with
    screen coordiantes $\alpha=\SI{-3.55945}{\admmass}$, $\beta=0$. The geodesic
    is integrated backwards in both the ingoing (solid blue line) and outgoing
    (red dashed line) Kerr-Schild coordiantes with black hole spin $a =
    \SI{0.7}{\admmass}$. Both integrations identify that the photon comes from
    the horizon, but one does so in a regular fashion (red dashed line), and the
    blue one takes an infinite amount of time (solid blue line). The filled
    circle indicates when the photon crosses the horizon. In the inset, we show
    the trajectories of the photons in their own evaluated coordiante systems.}
  \label{fig:kerr-phi}
\end{figure}

Figure~\ref{fig:kerr-phi} shows the constraint violation
($\lvert k^{\mu} k_{\mu} \rvert$, where $k^{\mu} $ is $ \d x^{\mu} \slash \d \lambda$) for a photon
integrated in ingoing or outgoing Kerr-Schild coordinates. Note that the
spinning case differs from the Schwarzschild one in the important fact that it
is not symmetric with respect to time reversal. Therefore, the geodesic obtained
when integrating photons towards the source is not the correct one to use for
radiation transfer. So, here we always integrate backwards in time with momentum
pointing away from the horizon. The figure shows that the integration is
well-behaved only for outgoing coordinates, in which case we can reconstruct the
entirety of the geodesic without running into numerical problems. When we use
ingoing Kerr-Schild coordinates, the constraint violation diverges. Photons that
fall into the horizon in the ingoing metric spend an infinite amount of time
orbiting the black hole in the outgoing ones. In doing this, they accumulate
numerical error for the same reason highlighted in
Section~\ref{sec:schwarzschild}. One possible remedy implemented by previous
studies is to impose a boundary condition so that the integration is stopped at
a distance that is larger than the distance at which the photon orbits, and an
alternative approach is presented was Section~\ref{sec:poss-work-integr}
(integrating with respect to the coordinate time). We find here the same
conclusion we found in the previous Section: using an optimal set of coordinates
results in higher accuracy and significantly higher performance, as it was shown
in the bottom panel of Figure~\ref{fig:kt-schw}, where the entire solution was
obtained with a handful of steps.

While some quantities depend critically on the gauge, the difference in
performance between the two coordinate systems has marginal effects on
coordinate-independent observables. Analytically, quantities that are obtained
through gauge-independent processes do not depend on the choice of coordinates,
but this is not necessarily true numerically because of the numerical error and
ad-hoc fixes or termination conditions. The accumulated numerical error near the
horizon can change results: Figure~\ref{fig:kt-schw} and
Figure~\ref{fig:kerr-phi} show that the violation of the constraint ($k^{\mu}
k_{\mu} = 0$, the null condition for the geodesic) explodes, meaning that the
integration becomes less and less accurate. In addition to that, arbitrary
termination conditions truncate prematurely the geodesics. In practice, in the
case of ill-suited coordinate systems, a ray will orbit the horizon an infinite
amount of times but the contribution to the specific intensity is suppressed in
the process.\footnote{If $I_{\nu}$ is the specific intensity $\d I_{\nu} \slash
\d \lambda \propto \upsilon^{-1} $, where $\upsilon$ is the redshift
factor~\cite{Younsi2012}. Unless one designed a pathological fluid
configuration, $\upsilon$ diverges with $k^{t}$ in ingoing Kerr-Schild
coordinates, so the contribution to the specific intensity vanishes as the
photon orbits the horizon.} Hence, the missing amount of the flux compared to
case with coordinates that are better suited for ray tracing is minimal.

\begin{figure}[htbp]
  \centering
  \includegraphics[]{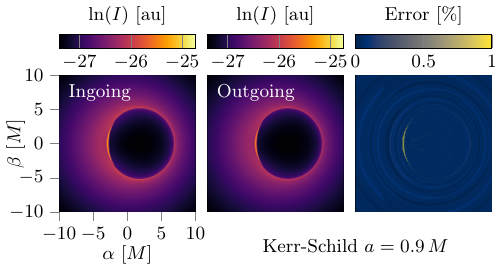}
  \caption{Measured intensity from a stationary fluid configuration emitting
    with thermal bremsstrahlung on different coordinate systems. The camera is
    on the equatorial plane ($\cal{i} = 90{}^{\circ}, \cal{j} = 0{}^{\circ}$) at a
    Euclidean distance $d$ of \SI{1e3}{\admmass}. Intensity is measured in
    arbitrary units (au). Error (right panel) is computed as the relative
    difference between the two solutions generated integrating the rays backward
    in time in the ingoing (left panel) and outgoing (middle panel) Kerr-Schild
    coordinates. }%
  \label{fig:image-comp}
\end{figure}

In Figure~\ref{fig:image-comp}, we compare the image produced by performing
gauge-invariant radiation transport \emph{backward in time} in both ingoing and
outgoing Kerr-Schild coordinates with spin $a = 0.9$. The radiation transfer is
implemented as in Equations~(19) and~(20) in~\cite{Younsi2012}. For this image,
we consider a toy example with a stationary fluid with four-velocity $(1 \slash
\sqrt{-g_{tt}}, 0, 0, 0)$ distributed with density that goes as $1\slash R$,
where $R$ is the Euclidean distance from the center of the black hole. The fluid
has a fixed temperature, and we assume its emissivity is thermal
bremsstrahlung~\cite{Rybicki1986}. This fluid configuration is not realistic but
similar results are obtained with other setups. Figure~\ref{fig:image-comp}
shows the resulting image and shows that the fractional difference is small. To
reduce the difference between the two images one has to move the integration
termination criterion closer and closer to the real horizon, and move the final
coordinate time to larger and larger values in the past. Due to the numerical
errors described in Section~\ref{sec:schwarzschild} and Section~\ref{sec:kerr},
in practice, it is impossible to obtain perfect convergence with
finite-precision codes.

A second quantity that is often studied in the literature is the black-hole
shadow (also known as the critical curve~\cite{Gralla2019}). These computations
of are also not much affected by the choice of coordinates because they rely on
binary identification (whether the photon comes from the horizon). Even if
integrations in ingoing coordinates collect significant errors and cannot be
completed, most stopping criteria will correctly identify whether the photon
originated from the horizon. For this reason, the computation will not be
affected by the large errors.

In conclusion, if one has access to an analytical spacetime and fluid model, we
recommend using coordinates that are adapted to the problem of ray tracing. In
the other cases, gauge-invariant calculations will still produce robust results.

\section{Conclusions}%
\label{sec:conclusions}

Ray tracing is a fundamental tool for our understanding of the observational
appearance of black holes. In this paper, we discussed how general relativistic
ray tracing is dependent on the adopted chart. In
Section~\ref{sec:schwarzschild}, we showed that in general the process is not
time reversible and integrating the geodesic equations towards the source
forward and backward in time lead to different results. This is against our
common intuition, according to which, there is only one light ray that connects
our eyes to a given object. We also discussed properties of coordinates, and
showed that charts that are designed to facilitate the flow of information into
horizons are not optimal choices for ray tracing. Using the best set of
coordinates results in significantly higher accuracy and performance. Hence,
we recommend to use suitable coordinates for those studies that use analytical
spacetimes and matter configurations. As shown in
Section~\ref{sec:poss-work-integr}, integrating in coordinate time is a good
solution for other cases (e.g., in GRMHD simulations). In
Section~\ref{sec:kerr}, we discussed how some quantities depend on the
coordinates, and showed that for gauge-independent observables (like black-hole
images obtained with radiation transfer) the numerical problems can lead to
small errors.

\begin{acknowledgments}
  We wish to thank Sam Gralla, Dimitrios Psaltis, Aniket Sharma, and Erik Wessel
  for useful discussions. We thank Pierre Christian for sharing a modified
  submodule of \texttt{FANTASY}~\cite{Christian2020} to compute the Christoffel
  symbols of Kerr-Schild spacetimes that we used to test our implementation.
  This work was in part supported by NSF Grants PHY-1912619, and PHY-2145421, as
  well as NASA Grant 80NSSC20K1542 to the University of Arizona, and a Frontera
  Fellowship by the Texas Advanced Computing Center
  (TACC). Frontera~\cite{Frontera2020} is funded by NSF grant
  OAC-1818253. Figure~\ref{fig:pen-kru} is based on a public TikZ code by Izaak
  Neutelings. This research made use of \texttt{GNU Parallel}~\cite{parallel},
  \texttt{SciPy}~\cite{SciPy}, \texttt{NumPy}~\cite{NumPy},
  \texttt{dill}~\cite{dill}, \texttt{SymPy}~\cite{SymPy}, and
  \texttt{kuibit}~\cite{kuibit}. Calculations were performed on \texttt{aitken}
  at the NASA Advanced Supercomputing center and on \texttt{puma} at the
  University of Arizona.

\end{acknowledgments}

\appendix*

\section{Integrating the radial geodesic in the Schwarzschild spacetime}%
\label{sec:integrating-kappa}

In Section~\ref{sec:schwarzschild}, we discussed the properties of the radial
geodesics in Schwarzschild spacetimes by looking at the null cones. Here, we
provide a full integration of the geodesic equation for outgoing solutions in
affine parameter and in ingoing Kerr-Schild coordinates.

Equation~\eqref{eq:two-null-solutions-integrated} provides the relationship
between $x$ and $t$ for null outgoing geodesics, let us compute $x(\lambda)$ and
$t(\lambda)$, where $\lambda$ is the affine parameter. Let us define $\kappa = \d \lambda \slash \d t$, from
the geodesic equation~\eqref{eq:geo-eq-chri}, we have that (as long as $\kappa$ is
finite and non-zero)
\begin{equation}
  \frac{\d \kappa}{\d t} = \kappa \Gamma^{t}_{\alpha\beta} \frac{\d {x}^{\alpha}}{\d t} \frac{\d {x}^{\beta}}{\d t}\,.
\end{equation}
The Christoffel symbols for metric~\eqref{eq:schwarzschild-kerr-schild} are
\begin{align*}
  \Gamma^{t}_{tt} &= \frac{2 M^{2}}{x^{3}}\,, \\
  \Gamma^{t}_{tx} &= \Gamma^{t}_{xt} = \frac{M (2 M + x)}{x^{3}}\,, \\
  \Gamma^{t}_{xx} &= \frac{2M (M + x)}{x^{3}}\,.
\end{align*}
From Equation~\eqref{eq:two-null-solutions}, we have that
$\dot{x} = (x - 2M)\slash (x + 2M)$, so we can find the differential equation
\begin{equation}
  \frac{1}{\kappa} \frac{\d \kappa}{\d x} = \frac{4M}{(x - 2 M)(x + 2M)}\,.
\end{equation}
We can solve this equation with separation of variables:
\begin{equation}
  \ln \kappa = \ln{(x - 2 M)} - \ln{(x + 2 M)} + K\,,
\end{equation}
with $K$ constant of integration, which we can fix to zero by assuming that
$\kappa = 1$ for $x \to +\infty$. Therefore, we have that
\begin{equation}
  \kappa(x) = \frac{x - 2M}{x + 2M}\,.
\end{equation}
From Equation~\eqref{eq:two-null-solutions-fixed}, we recognize $\d x \slash \d
t$ on the left-hand-size of this last equation, which, coupled with the definition of $\kappa$, leads
to
\begin{equation}
  \frac{\d \lambda}{\d t} = \frac{\d x}{\d t}\,.
\end{equation}
Integration of this equation shows that $x$ is an affine parameter. We conclude
that
\begin{equation}
  x(\lambda) = \lambda + L\,,
\end{equation}
where $L$ is a constant that can be determined by demanding that $x(0)$ is the
location of the camera. Plugging in Equation~\eqref{eq:two-null-solutions}, we
find the expression for $t(\lambda)$.

\bibliography{ideal_coordinates_ray_tracing,adsbib}

\end{document}